\documentclass[prl,aps,showpacs,superscriptaddress,twocolumn]{revtex4}
\usepackage{amsmath}
\usepackage{amssymb}
\usepackage{color}
\usepackage {graphicx}

\newcommand{\KFA}{KFe$_2$As$_2$}

\newcommand{\AFS}{AFe$_2$Se$_2$}

\newcommand{\beq}{\begin{equation}}
\newcommand{\eeq}{\end{equation}}
\newcommand{\bea}{\begin{eqnarray}}
\newcommand{\eea}{\end{eqnarray}}

\begin{document}
\title{Evolution of superconductivity in Fe-based systems with doping}
\author{S.~Maiti}
 \affiliation{Department of Physics, University of Wisconsin, Madison, Wisconsin 53706, USA}
\author{M.M.~Korshunov}
 \affiliation{Department of Physics, University of Florida, Gainesville, Florida 32611, USA}
 \affiliation{L.V. Kirensky Institute of Physics, Siberian Branch of Russian Academy of Sciences, 660036 Krasnoyarsk, Russia}
\author{T.A.~Maier}
 \affiliation{Computer Science and Mathematics Division and Center for Nanophase Materials Sciences, Oak Ridge National Lab, Oak Ridge, TN 37831, USA}
\author{P.J.~Hirschfeld}
 \affiliation{Department of Physics, University of Florida, Gainesville, Florida 32611, USA}
\author{A.V.~Chubukov}
 \affiliation{Department of Physics, University of Wisconsin, Madison, Wisconsin 53706, USA}

\date{\today}


\begin{abstract}
We study the symmetry and the structure of the gap in
 Fe-based superconductors by decomposing the pairing interaction
obtained in the RPA into $s$- and $d$-wave
 components and into contributions from scattering between
different Fermi surfaces. We show that each interaction  is well
approximated by the lowest angular harmonics and use this
simplification to analyze the origin of the attraction in $s^\pm$
and $d_{x^2-y^2}$ channels, the competition between $s$- and
$d$-wave solutions, and the origin of superconductivity in heavily
doped systems, when only electron or only hole pockets  are
present.
\end{abstract}
\maketitle
{\it Introduction.} The symmetry and structure of the
superconducting gap in Fe-based superconductors (FeSC), and their
evolution and possible change with doping are currently subjects of
intensive debates in the condensed matter community. The vast
majority of researchers believe that superconductivity in FeSC is of
electronic origin and results from the screened Coulomb interaction,
enhanced at particular momenta due to strong magnetic fluctuations.
 In FeSC, the Fermi surface (FS) has multiple sheets due to
hybridization of all five $d$-orbitals of Fe, and the interactions
between low-energy fermions are a complex mixture of contributions
from intra - and inter-orbital interactions. In this situation, both
$s$-wave and non-$s$-wave pairing are possible, and
 can be either conventional or extended, with $\pi$ phase shifts
between  FSs~\cite{wen}

Previous theoretical works on
 FeSCs with hole and
electron pockets  have
shown~\cite{tom_09,kuroki,arita,graser,peter,peter_2,Rthomale,Thomale,dhl_10,zlatko,Andrey,saurabh}
that the $s$-wave pairing channel (for sign-changing $s^\pm$ gap)
 is generally the most attractive,
although the $d$-wave channel is a strong competitor. $s$-wave gap
symmetry is consistent with ARPES data, which detected only a small
variation of the gap along the hole FSs, centered at $(0,0)$, and as
such ruled out $d$-wave gap symmetry
  However, for the recently
discovered heavily electron-doped \AFS
(A=K, Rb, Cs)~\cite{exp:AFESE}, in which only electron FSs remain according to ARPES~\cite{exp:AFESE_ARPES}, RPA and functional RG (fRG)
studies found that the leading pairing instability is now in the $d$-wave channel.~\cite{dhl_AFESE,graser_11} $d$-wave pairing was also found in an
fRG study of heavily hole-doped \KFA~\cite{KFeAs_fRG}, in which
only hole FSs are present~\cite{KFeAs_ARPES_QO}. For this
material, various experimental probes~\cite{KFeAs_exp_nodal}
indicate the presence of gap nodes,
 consistent with a $d$-wave gap symmetry.

In this communication, we analyze the competition between $s$-
and $d$-wave pairings in doped FeSCs, the origin of attraction at
small and large dopings,
  and the structure of
$s$- and $d$-wave gaps at various dopings. We argue that the
pairing mechanisms at small and large dopings are qualitatively
different and that the $d$-wave state at large hole doping is a
different eigenstate from the one
that competes with $s$-wave at smaller dopings.

We assume, as in earlier works, that FeSCs can be treated as
itinerant systems, and that the pairing interaction is enhanced by
  spin fluctuations (SF). In the band description adopted here,
 the electronic structure at low energies is obtained by
hybridization of all five Fe $d$-orbitals and in electron-doped
FeSCs consists of two cylindrical hole FSs
  $h_1$ and $h_2$, centered at $(0,0)$,
  and two cylindrical electron FSs $e_1$ and $e_2$,
 centered at $(\pi,0)$ and $(0,\pi)$, respectively, in the
1-Fe zone. For hole-doped FeSCs, there exists an additional
cylindrical hole FS $h_3$ centered at $(\pi,\pi)$. In such a description,
interactions are dressed by matrix elements associated with the hybridization of
orbitals, and depend on the
 angles
 along the FSs. 

{\it The method.} The input for our analysis is the band model
with the interactions between the particles on the FSs $\Gamma (\mathbf{k}_F, -\mathbf{k}_F; \mathbf{k}'_F, -\mathbf{k}'_F) \equiv \Gamma (\mathbf{k}_F,
\mathbf{k}'_F)$. The interactions are obtained numerically in the RPA
SF formalism starting from the 5-orbital  model
\cite{graser} with intra- and inter-orbital hoppings
and  local density-density and exchange interactions $U$, $U'$,
$J$, and $J'$.
We show that, in the band basis,
 each interaction component  $\Gamma_{ij} (\mathbf{k}_F, \mathbf{k}'_F)$ is well approximated by the leading angular harmonics (LAH) in $s$-wave and $d_{x^2-y^2}$-wave channels 
 (similar to the approximation of the  $d_{x^2-y^2}$ gap by $\cos 2 \theta$ in the cuprates),
 and use the LAH
approximation (LAHA) to reduce $s$-wave and $d$-wave gap equations
to either $4 \times 4$ or $5 \times 5$ sets which can be easily solved
and analyzed. This allows us to go a step further than previous
works and understand the pairing mechanism at different dopings, the
origin of the transition from $s$-wave to $d$-wave instability, the
role of the SF component of the interaction, and the stability of
$s$-wave and $d$-wave gap structures with respect to the variation
of parameters in the gap equations. For simplicity, we assume in LAHA
 that all FSs are circular, with the same density of states $N_F$. The results change only a little if we use the actual lattice fermionic dispersion.

 The application of LAHA for FeSCs requires some care, as electron FSs are
centered at $(0,\pi)$ and $(\pi,0)$ points, which are not $k_x
\leftrightarrow \pm k_y$ symmetric. As a result, some of the
$s$-wave gap functions, like $\cos k_x + \cos k_y$ behave as $\pm
\cos 2\theta$ along the electron FSs, while some of the $d$-wave gap
functions like $\cos k_x - \cos k_y$ are approximated on these FSs
by constants of opposite sign. With this in mind, we treated the
angle-independent and $\cos 2\phi$ terms on equal footings in both
$s$-wave and $d$-wave components of the interactions. A simple
 analysis then shows that LAHA consistent with the FS geometry of
FeSCs approximates the $s$ and $d_{x^2-y^2}$ components of ${\bar
\Gamma_{i,j}} = N_F \Gamma_{ij}$ as
\bea \label{eq:interactions}
 \bar\Gamma_{h_ih_j} &=& u_{h_ih_j}+\tilde{u}_{h_ih_j} \cos2\phi_i \cos2\phi_j \\
 \bar\Gamma_{h_ie_1} &=& u_{h_ie} (1 + 2\alpha_{h_ie} ~\cos2\theta_1)\nonumber\\ &&+\tilde{u}_{h_ie} (1+2{\tilde \alpha}_{h_ie} \cos2\theta_1) \cos2\phi_i \nonumber\\
 \bar\Gamma_{e_1e_1} &=& u_{ee} \left(1 + 2 \alpha_{ee}(\cos2\theta_1 + \cos2\theta_2)+\right. \nonumber\\
 && \left. 4\beta_{ee} \cos2\theta_1 \cos2\theta_2\right) + {\tilde u}_{ee} \left(1 +\right. \nonumber \\
 && \left. 2 {\tilde \alpha}_{ee} (\cos2\theta_1+ \cos2\theta_2)+4 {\tilde \beta}_{ee} \cos2\theta_1 \cos2\theta_2\right) \nonumber
\eea
where
$u_{ij}$ and ${\tilde u}_{ij}$ are dimensionless interactions in $s$-wave and $d$-wave channels, respectively, and $\phi_i$ and $\theta_i$ label the angles along the hole and electron FSs, measured from the $k_x$-axis.
Interactions involving other electron FSs are obtained by
transformations consistent with $s$-wave or $d$-wave symmetry.

We use Eq.~(\ref{eq:interactions}) to fit the RPA interaction
$\Gamma_{ij}$ by LAHA and substitute the parameters extracted from
the fit into $s$-wave and $d$-wave BCS gap equations, which within
LAHA are $4 \times 4$ matrix equations for two hole and two electron
FSs and $5 \times 5$ when the additional hole FS is present. We find
the gap structure for the largest positive eigenvalue
$\lambda_{s,d}$ (if it exists) and then vary the parameters $u_{ij}$
by hand to understand what is the mechanism for the attraction. For
two hole and two electron FSs the generic gap structure is
\bea
&&\Delta^{s}_{h_1} (\phi) =\Delta^{s}_{h_1},~~ \Delta^{s}_{h_2} (\phi) =\Delta^{s}_{h_2}  \label{5} \\
&&\Delta^{s}_{e_1} (\theta) =\Delta^{s}_{e} + {\bar \Delta}^{s}_e \cos 2
\theta, ~~
\Delta^{s}_{e_2} (\theta) =\Delta^{s}_{e} - {\bar \Delta}^{s}_e \cos 2 \theta \nonumber\\
&&\Delta^{d}_{h_1} (\phi) = \Delta^{d}_{h_1} \cos 2 \phi,~~\Delta^{d}_{h_2} (\phi) =  \Delta^{d}_{h_2} \cos 2 \phi\nonumber \\
&&\Delta^{d}_{e_1} (\theta) =\Delta^{d}_{e} + {\bar \Delta}^{d}_e
\cos 2 \theta, ~\Delta^{d}_{e_2} (\theta) = -\Delta^{d}_{e} +
{\bar \Delta}^{d}_e \cos 2 \theta \nonumber
\eea
and for five FSs we add one more $\Delta^{s,d}_{h_3} (\phi) =
\Delta^{s,d}_{h_3}$.

\begin{figure}[htp]
$\begin{array}{cc}
\includegraphics[width=0.45\columnwidth]{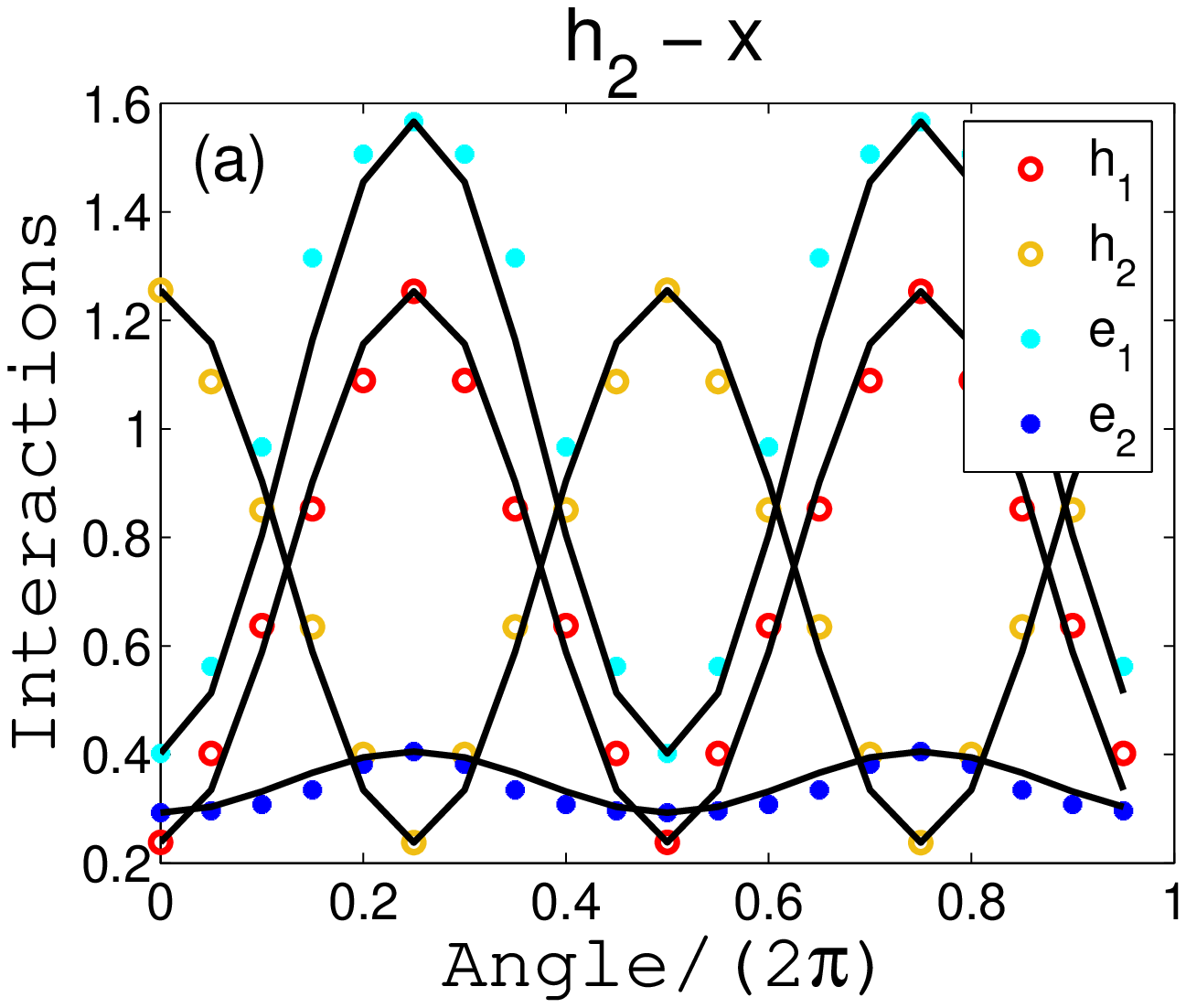}&
\includegraphics[width=0.45\columnwidth]{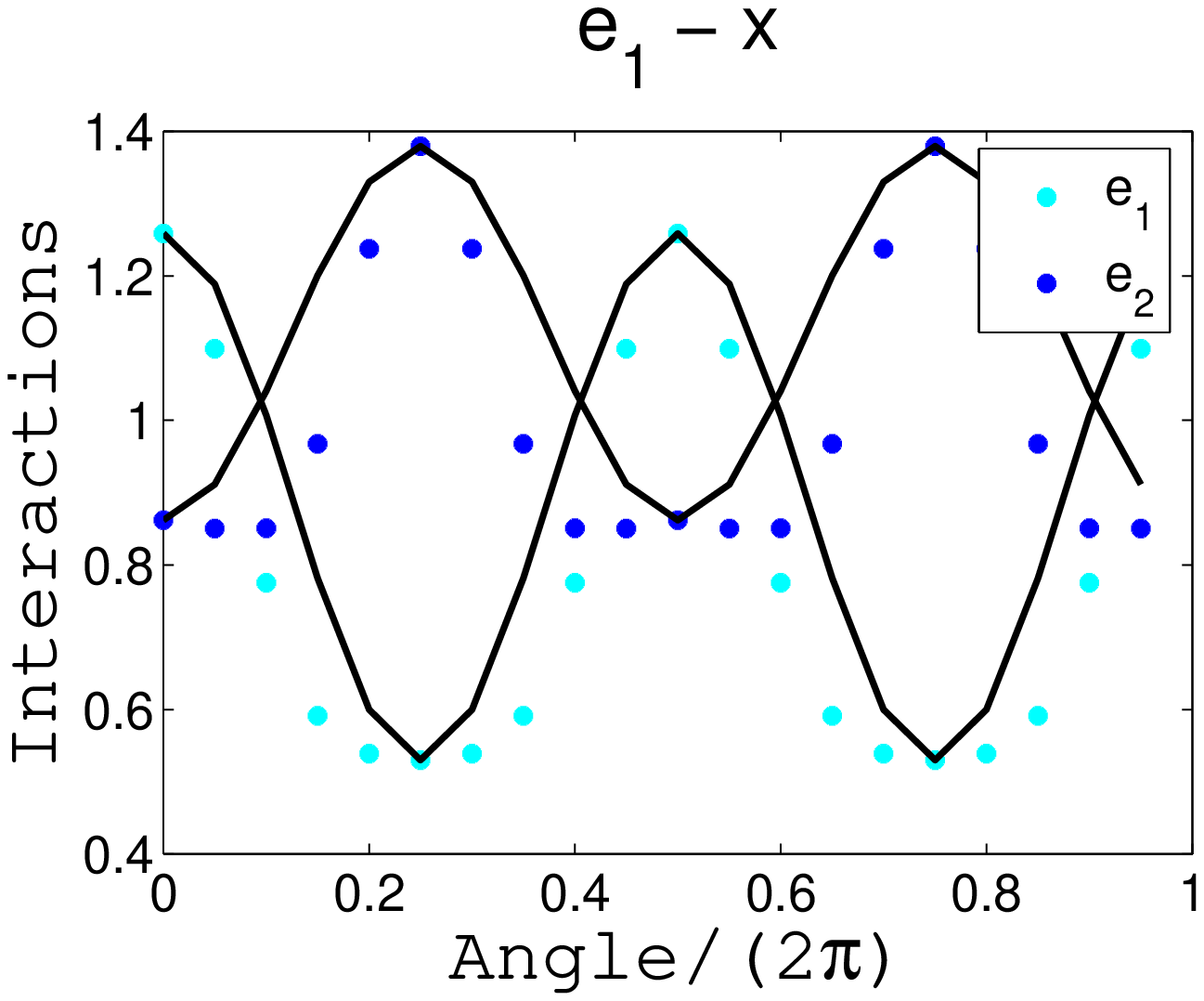}\\
\includegraphics[width=0.45\columnwidth]{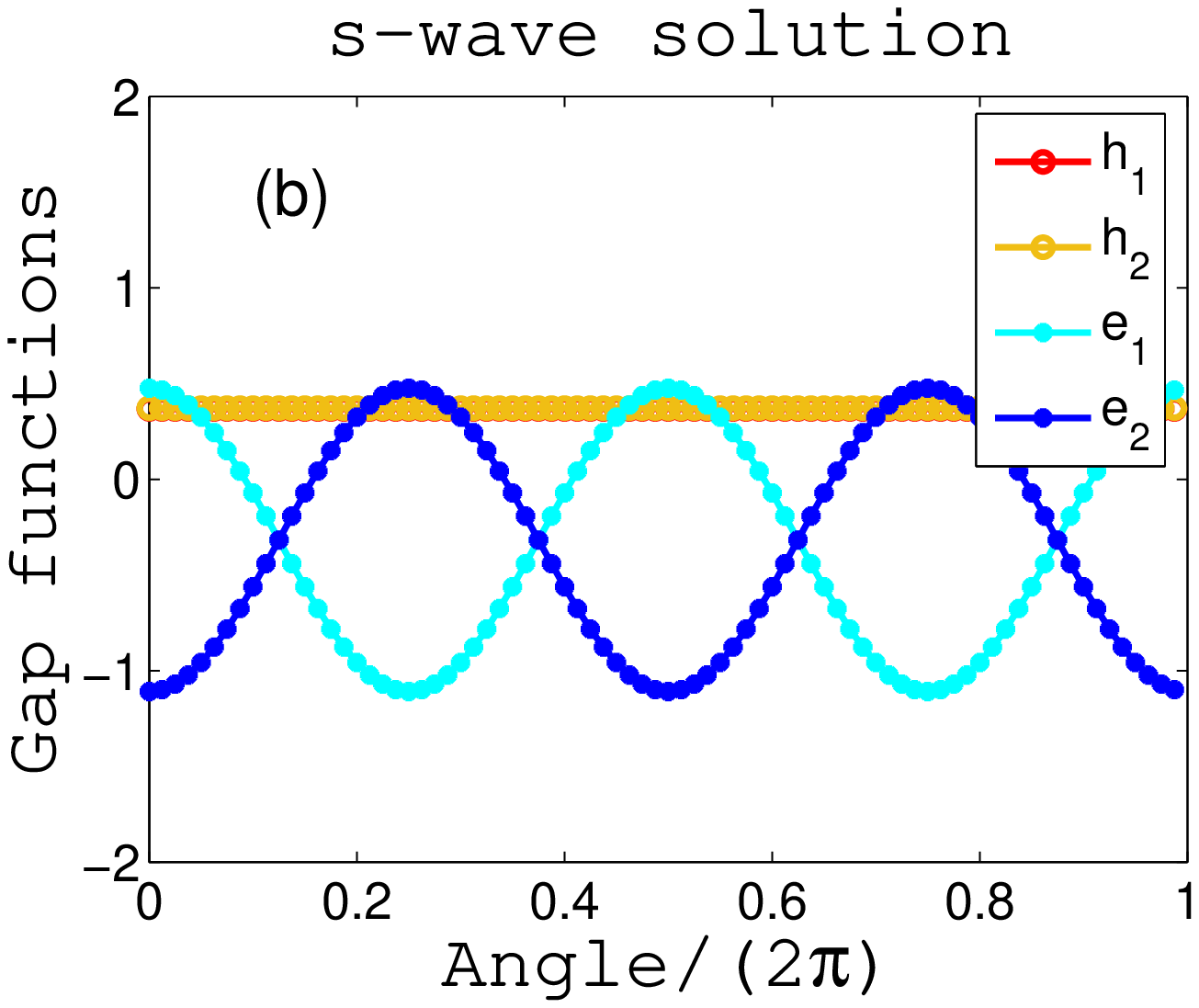}&
\includegraphics[width=0.45\columnwidth]{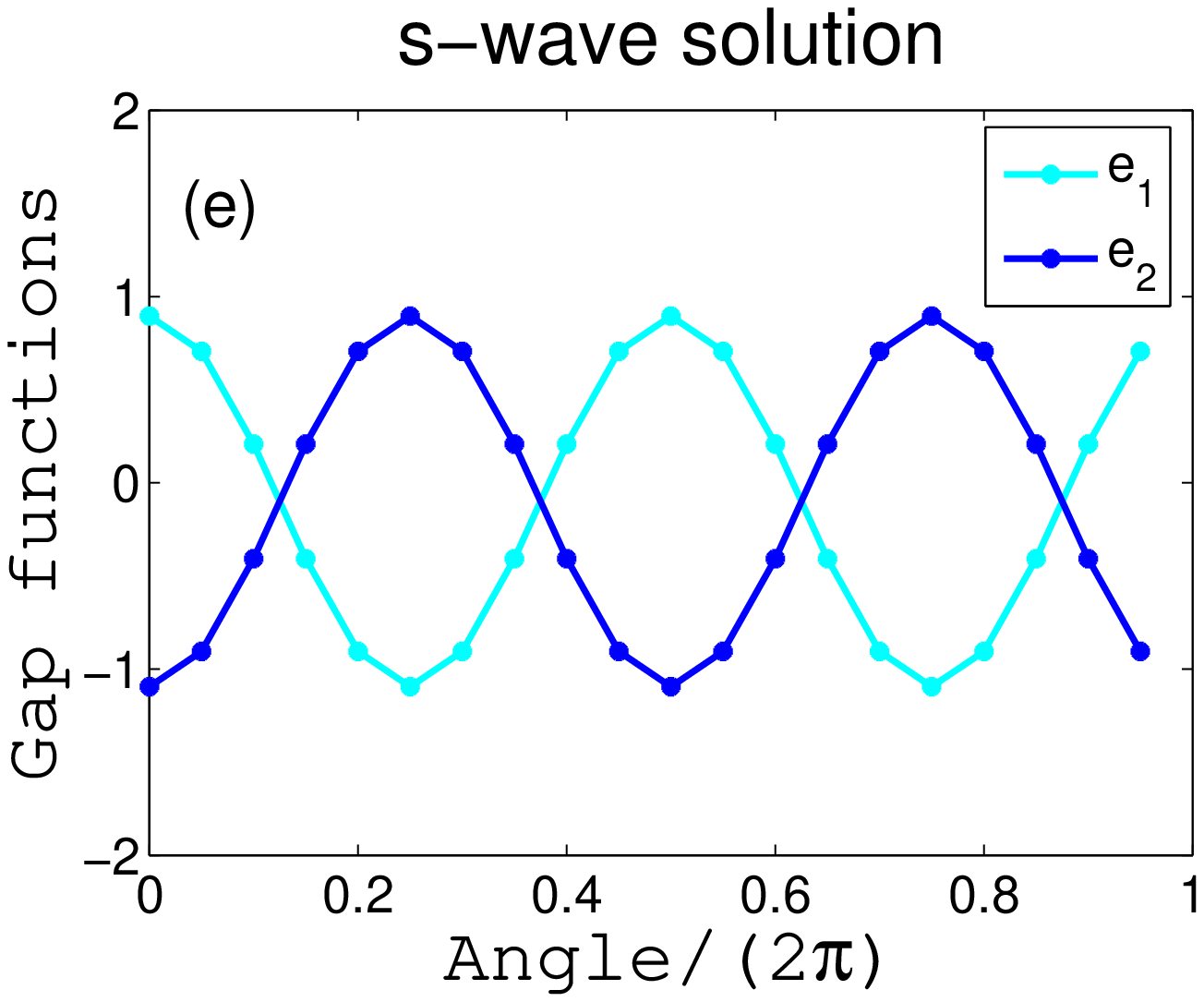}\\
\includegraphics[width=0.45\columnwidth]{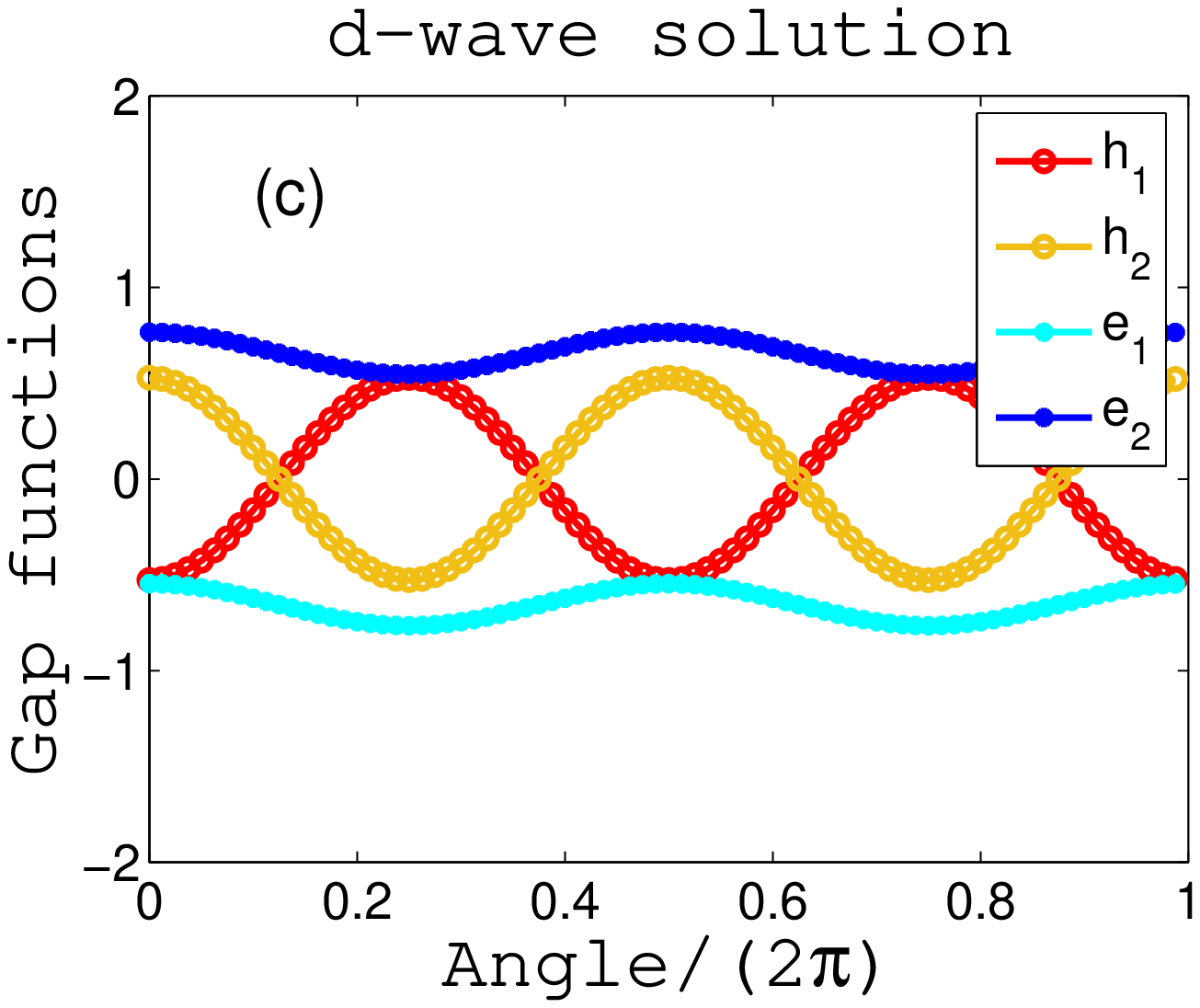}&
\includegraphics[width=0.45\columnwidth]{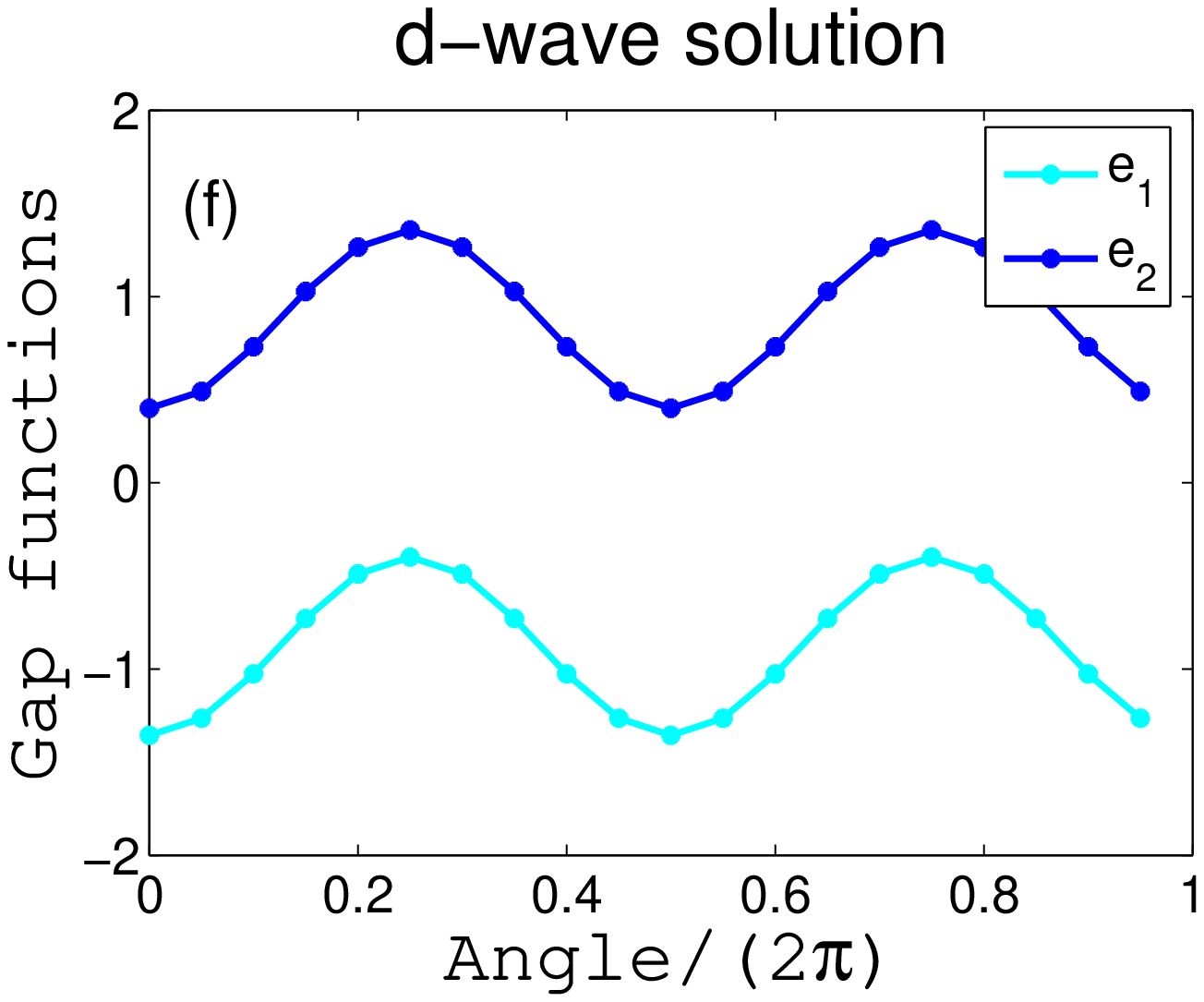}
\end{array}$
\caption{\label{fig:1} Electron doping.
(a-c) Representative LAHA fit of the
interactions $\Gamma(\mathbf{k}_F, \mathbf{k}'_F)$ and $s$- and $d$-wave gap functions for the case of  two very tiny hole
pockets. $\mathbf{k}_F$ is taken to be along $x$ on the $h_2$ FS,
while $\mathbf{k}'_F$ is varied along each of FSs.  The symbols
represent the RPA interactions computed numerically for the 5-band
model \cite{graser} using the LDA band structure \cite{ref:Cao},
the black lines are the fits using Eq. \ref{eq:interactions}.
Angle is measured relative to $k_x$.
 (d)-(f) are the same as (a)-(c) but for stronger electron doping,
  where there are no hole pockets. The
parameters are presented in \cite{comm_2}.}
\end{figure}

\begin{figure}[htp]
$\begin{array}{ccc}
\includegraphics[width=0.45\columnwidth]{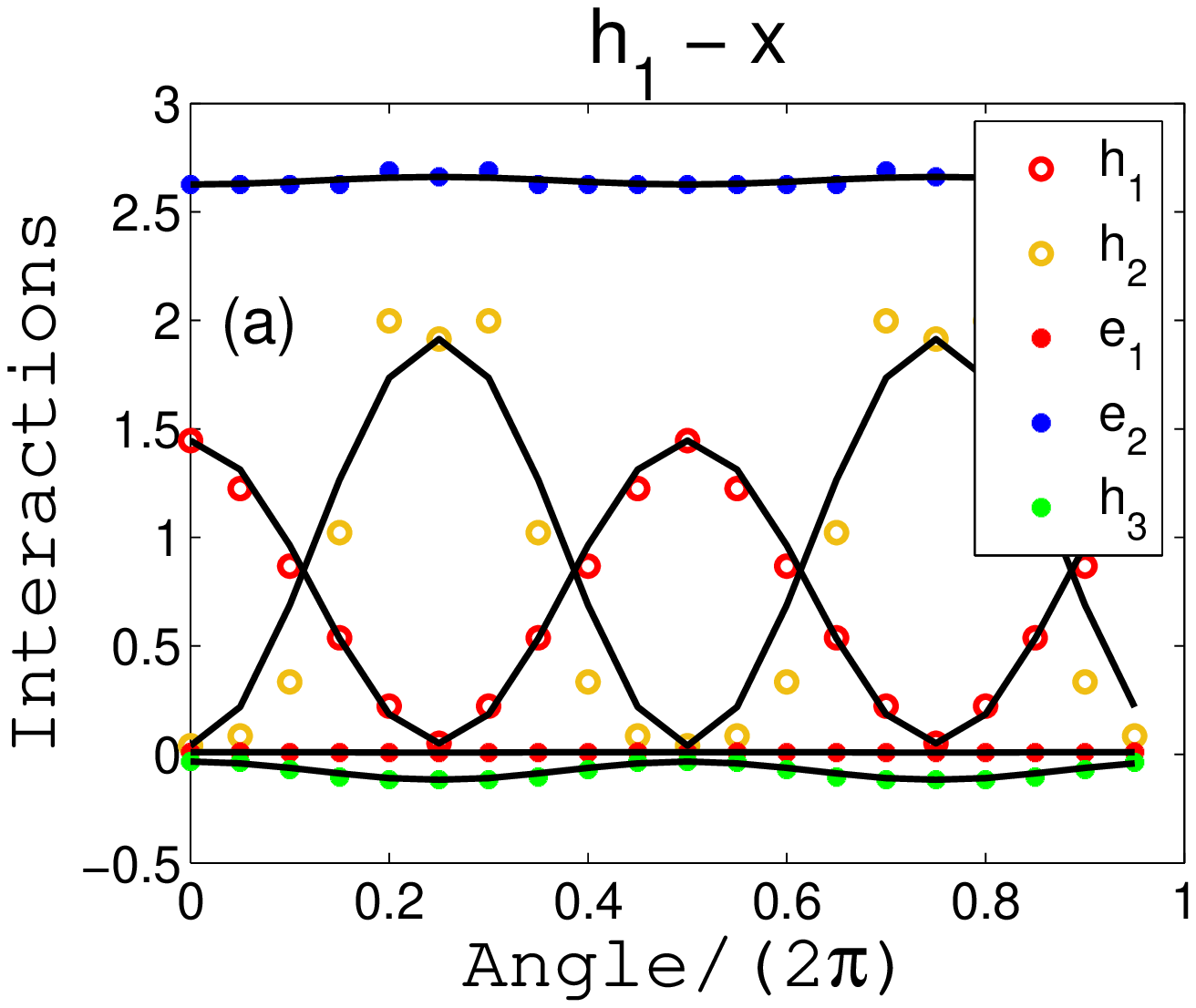}&
\includegraphics[width=0.45\columnwidth]{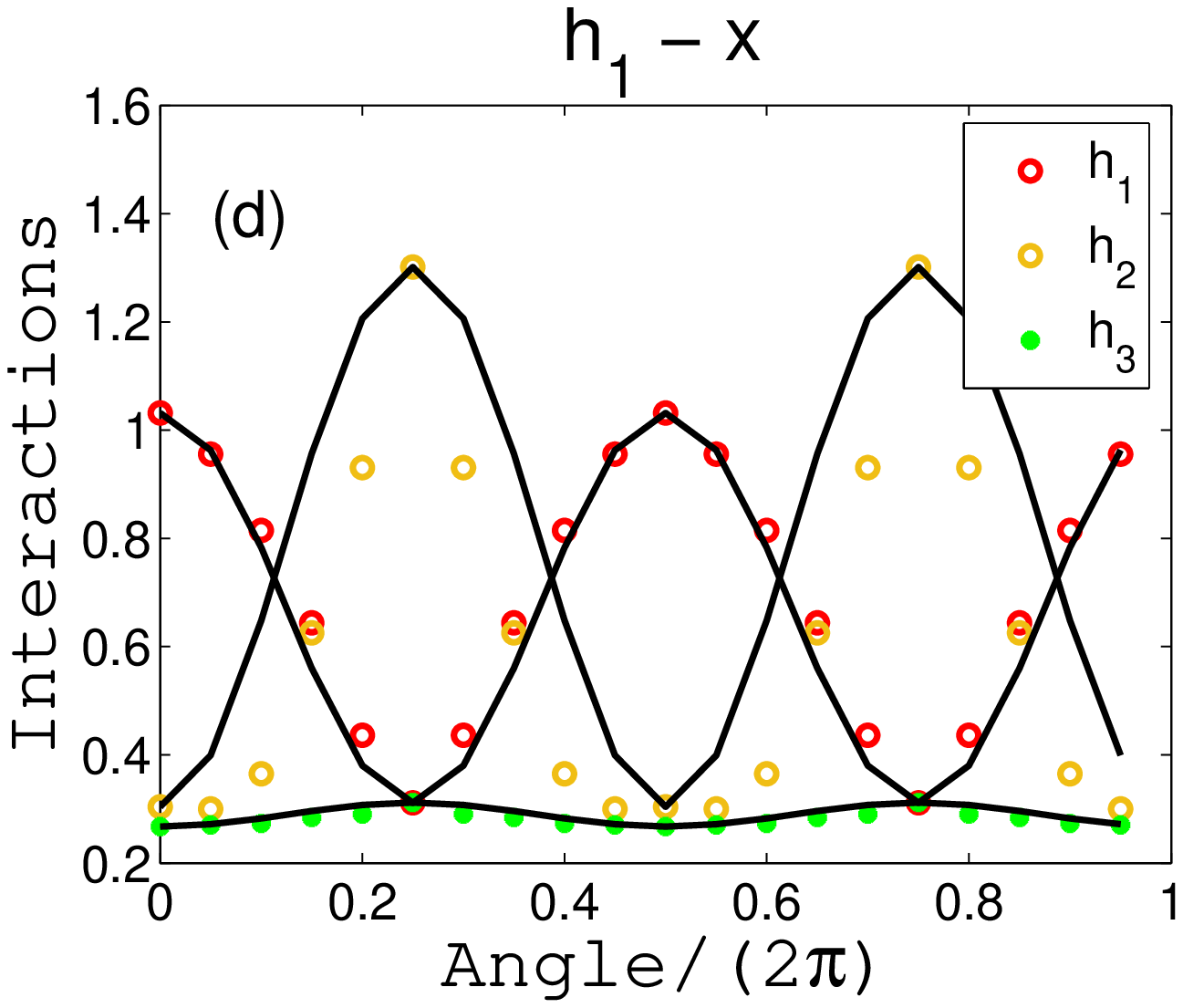} \\
\includegraphics[width=0.45\columnwidth]{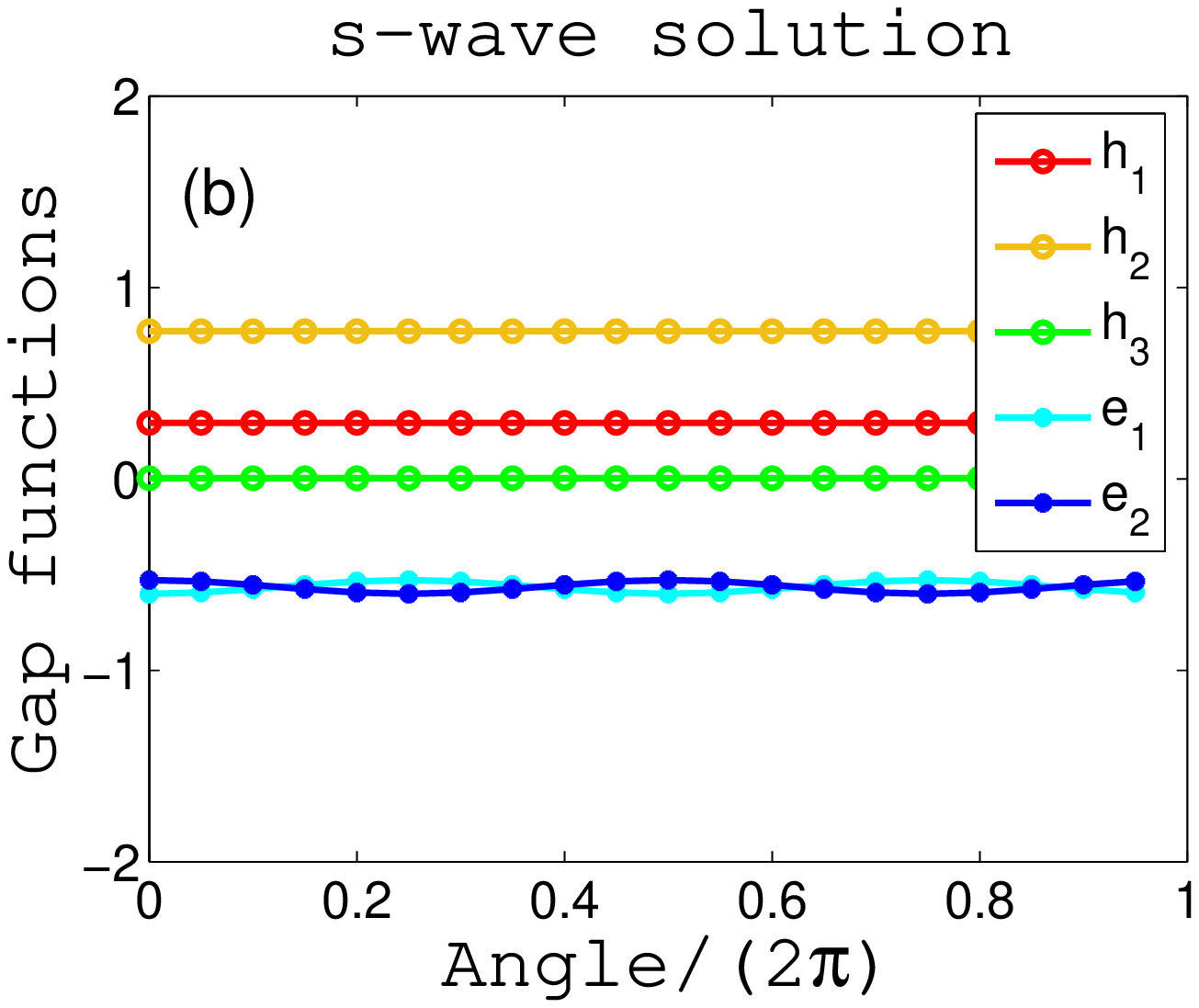}&
\includegraphics[width=0.45\columnwidth]{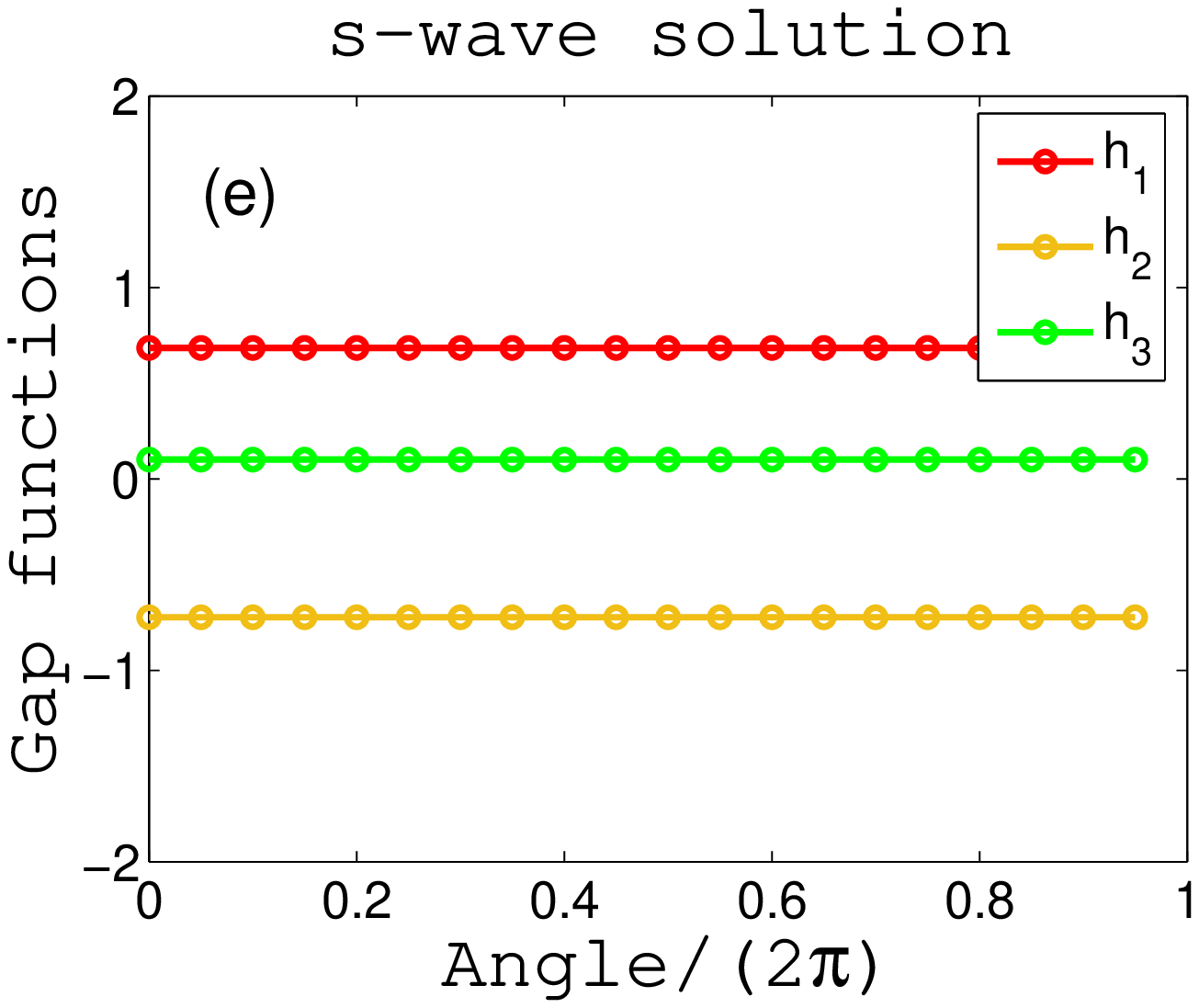}\\
\includegraphics[width=0.45\columnwidth]{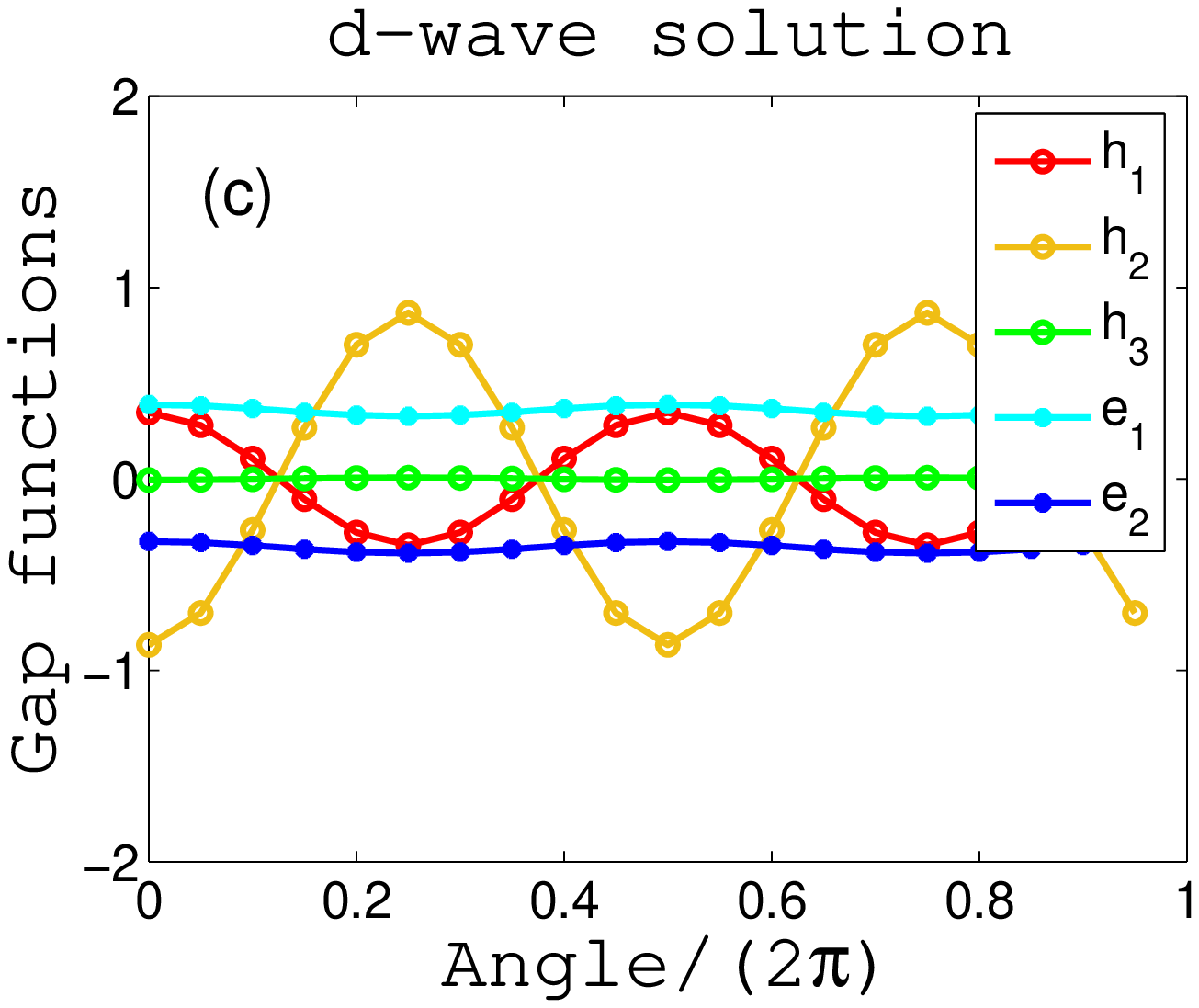}&
\includegraphics[width=0.45\columnwidth]{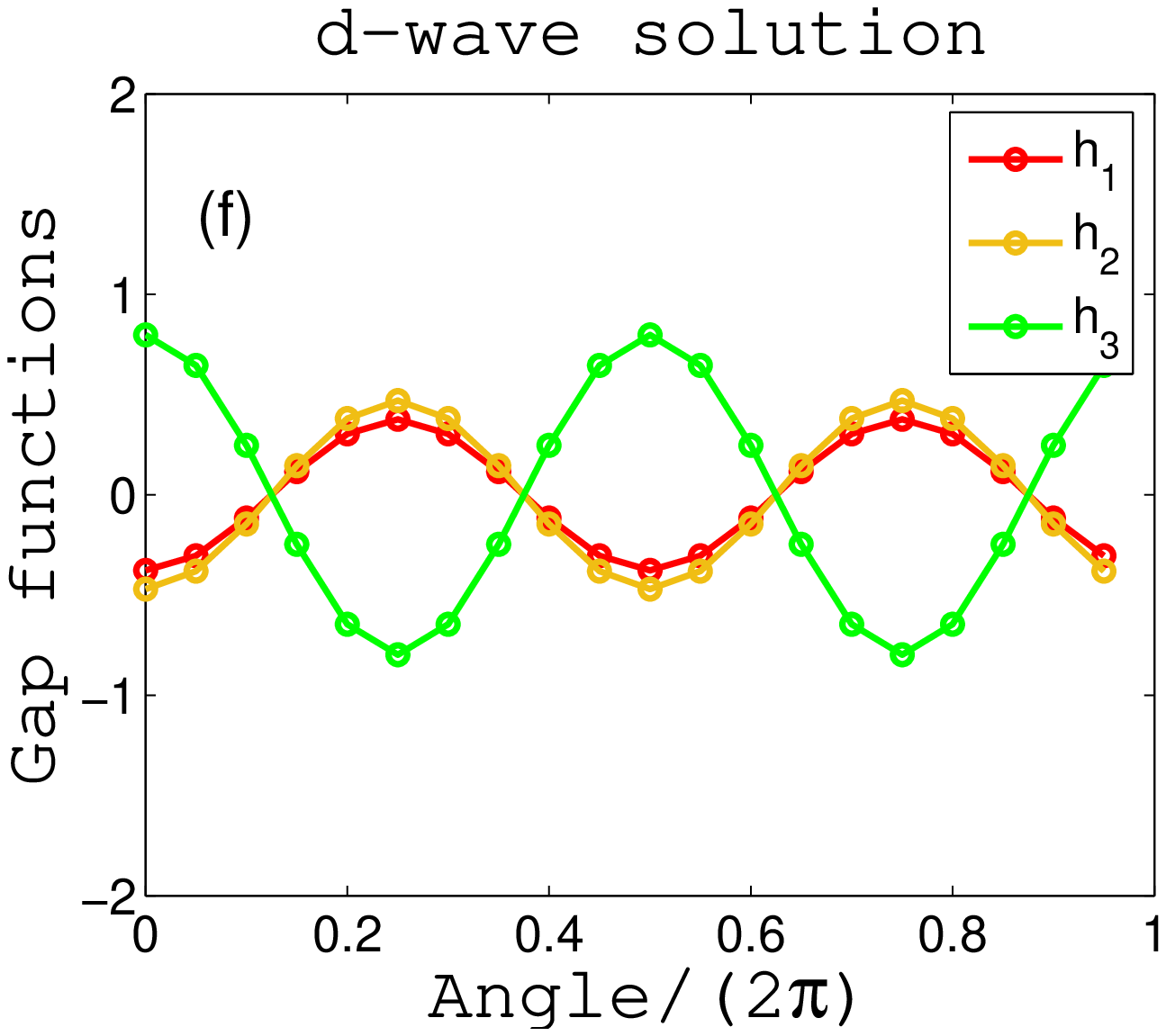}
\end{array}$
\caption{\label{fig:2} The same as in Fig.~\ref{fig:1}, but for
hole doping (3 hole FSs).  Panels (a)-(c) are for
 the case of
  tiny electron pockets, (d)-(f) are for stronger hole doping,
 when there are no electron pockets. The parameters
 are presented in \cite{comm_2}.}
\end{figure}

In Figs.~\ref{fig:1} and \ref{fig:2} we compare LAHA with the full
RPA $\Gamma_{ij}(\mathbf{k}_F, \mathbf{k}'_F)$. The agreement is
remarkably good. We analyzed eight different sets of $U$, $U'$ and
$J$, and the agreement is equally good for all sets~\cite{comm_2}.
A very few disagreements are cured by adding $\cos 4\theta$
harmonics to LAHA. Some of the LAHA parameters extracted from the fit,
which we will need for comparisons, are shown in Tables~\ref{tab:1} and \ref{tab:2}.
\begin{table}[htp]
\caption{\label{tab:1} Some of the LAHA parameters  extracted from the fit in Fig.~\protect\ref{fig:1} for electron doping. Block (i) corresponds
to panels (a)-(c) (tiny hole pockets), block (ii)
corresponds to panels (d)-(f) (no hole pockets).}
\begin{ruledtabular}
\begin{tabular}{lccccccrccc}
& \multicolumn{6}{c}{(i)} & & \multicolumn{3}{c}{(ii)}\\
 \cline{2-7} \cline{9-11}
$s$-wave&$u_{h_1h_1}$&$u_{h_1e}$&$\alpha_{h_1 e}$&$u_{ee}$&$\alpha_{e e}$&$\lambda_s$& & $u_{e e}$&$\alpha_{e e}$&$\lambda_s$\\
&0.75&0.67&-0.19&0.88&0.1&0.21& & 0.84&0.09&-0.12\\
 \cline{2-7} \cline{9-11}
$d$-wave&$\tilde{u}_{h_1h_1}$&$\tilde{u}_{h_1e}$&$\tilde{\alpha}_{h_1 e}$&$\tilde{u}_{e e}$&$\tilde{\alpha}_{e e}$&$\lambda_d$& & $\tilde{u}_{e e}$&$\tilde{\alpha}_{e e}$&$\lambda_d$\\
&0.51&-0.32&-0.50&-0.05&0.9&0.35& & -0.04&0.88&0.13\\
\end{tabular}
\end{ruledtabular}
\end{table}

The cases of weak electron and hole dopings were solved
numerically within RPA in earlier works, and we verified that LAHA
results are very close to the full solutions. For brevity, we
present only the results for larger dopings, when one type of
pockets either almost or completely disappears. We will see that 
there are quite abrupt changes between the two regimes.

{\it Results and discussion.} We varied the magnitudes and angle
dependencies of the interactions by hand and checked what most
influences the value of $\lambda$ and the structure of the gap. We
found that some system properties are sensitive to the ratios of
the parameters, but some are quite universal.

For electron doping, parameter-sensitive properties include the gap
symmetry, since $\lambda_s$ and $\lambda_d$ remain comparable as
long as both hole and electron FSs are present (see Table
\ref{tab:1}), and the presence or absence of accidental nodes in the
$s$-wave gap, although for most of parameters
the gap does have nodes, as in Fig.~\ref{fig:1}(b). The universal
observation is that the driving force for attraction in both
$s$-wave {\it and} $d$-wave channels is the inter-pocket
electron-hole interaction ($u_{h_i e}$ and ${\tilde u}_{h_i e}$
terms), {\it no matter how small the hole pockets are}. When the SF
component of the interaction is large,
 $u_{h_i e}$ and ${\tilde u}_{h_i e}$ exceed the
hole-hole and electron-electron interactions. Then 
 $\lambda_{s,d}$ are positive already if we neglect
the $\cos 2\theta$ terms in (\ref{eq:interactions}) (for two equal
hole FSs the conditions are $u^2_{he} > u_{hh} u_{ee}$ and
$\tilde{u}^2_{he} > \tilde{u}_{hh} \tilde{u}_{ee}$). In this case,
the $\cos 2\theta$ terms in the $s$-wave and $d$-wave gaps scale
with the corresponding $\alpha_{he}$. For smaller SF component, when
$u^2_{he} < u_{hh} u_{ee}$ (the case considered in Fig.~\ref{fig:1}
and Table~\ref{tab:1}), the electron-hole interaction still
generates solutions with $\lambda_{s,d} >0$, only this time the gap
develops a stronger $\cos 2 \theta$ component, which effectively reduces $u_{ee}$.

The situation changes qualitatively once the hole pockets disappear
(Fig.~\ref{fig:1}(d)-(f)). We see from Table~\ref{tab:1} that
$\lambda_s$ is reduced, but $\lambda_d$ is enhanced, i.e., the
$d$-wave $T_c$ increases. Comparing the LAHA parameters for the two
dopings, we see the reason: once the hole pockets disappear, a
direct $d$-wave electron-electron interaction ${\tilde u}_{ee}$
becomes strong and attractive. To understand why this happens, we
note that $u_{ee}$ and ${\tilde u}_{ee}$ are symmetric and
antisymmetric combinations of intra-pocket and inter-pocket
electron-electron interactions: $u_{ee} = u_\mathrm{intra}^{ee} + u_\mathrm{inter}^{ee}$,
${\tilde u}_{ee} = u_\mathrm{intra}^{ee} - u_\mathrm{inter}^{ee}$. Both $u_\mathrm{inter}^{ee}$ and
$u_\mathrm{intra}^{ee}$ are positive (repulsive), hence $u_{ee} > 0$, but the
sign of ${\tilde u}_{ee}$ depends on the interplay between
$u_\mathrm{inter}^{ee}$ and $u_\mathrm{intra}^{ee}$.
 As long as
the hole FS is present, SF are peaked near $\mathbf{q}=(0,\pi)$ and
$(\pi,0)$, which are an equal distance from the relevant momenta
$\mathbf{q}=0$ for $u_\mathrm{intra}^{ee}$ and $\mathbf{q}=(\pi,\pi)$ for $u_\mathrm{inter}^{ee}$.
In this situation, $u_\mathrm{intra}^{ee}$ and $u_\mathrm{inter}^{ee}$
 remain close in magnitude,
and ${\tilde u}_{ee}$ is small.
 Once the hole pocket disappears, the peak in the RPA spin
susceptibility shifts towards $(\pi,\pi)$ ~\cite{graser_11} and
$u_\mathrm{inter}^{ee}$ increases more due to the SF component than
$u_\mathrm{intra}^{ee}$. A negative $u_\mathrm{intra}^{ee} -
u_\mathrm{inter}^{ee}$ then gives rise to a ``plus-minus'' gap on
the two electron FSs for the same reason that large $u_{he}$ gives
rise to a sign-changing gap between hole and electron FSs, (and
the interaction between hot spots in the cuprates gives rise to a
sign-changing gap in the hot regions). Such a gap changes sign
between electron pockets, which differ by $k_x \to k_y$ and
therefore has $d_{x^2-y^2}$ symmetry~\cite{dhl_AFESE,graser_11}.  Our gap functions in both $s$- and $d$-wave channels at large electron
doping are in good quantitative agreement with the full solution
of the RPA gap equation~\cite{graser_11} and with fRG
results~\cite{dhl_AFESE}.

\begin{table}[htp]
\caption{\label{tab:2} Some of LAHA parameters  extracted from the fit in Fig.~\protect\ref{fig:2} for hole doping. Block (i) corresponds to
panels (a)-(c) (tiny electron pockets), block (ii) corresponds
to panels (d)-(f) (no electron pockets).}
\begin{ruledtabular}
\begin{tabular}{lcccccrccccccc}
& \multicolumn{5}{c}{(i)} & & \multicolumn{5}{c}{(ii)}\\
 \cline{2-6} \cline{8-12}
$s$&$u_{h_1h_1}$&$u_{h_1e}$&$\alpha_{h_1 e}$&$u_{ee}$&$\lambda_s$& & $u_{h_1h_1}$&$u_{h_1h_2}$&$u_{h_1
h_3}$&$u_{h_3 h_3}$&$\lambda_s$\\
&0.75&1.36&0.08&1.40&1.8& & 0.67&0.8&0.29&1.37&0.13\\
\cline{2-6} \cline{8-12}
$d$&$\tilde{u}_{h_1h_1}$&$\tilde{u}_{h_1e}$&$\tilde{\alpha}_{h_1 e}$&$\tilde{u}_{e
e}$&$\lambda_d$& & $\tilde{u}_{h_1h_1}$&$\tilde{u}_{h_1 h_2}$&$\tilde{u}_{h_1 h_3}$&$\tilde{u}_{h_3 h_3}$&$\lambda_d$\\
&0.70&-1.32&0.0&1.45&1.2& & 0.36&-0.5&-0.02&-0.17&0.11\\
\end{tabular}
\end{ruledtabular}
\end{table}

Next we consider the case of hole doping. The LAHA fits to the cases
when electron FSs are small but still present and when only hole FSs
remain are shown in Fig.~\ref{fig:2}. The parameters extracted from
the fit are shown in Table~\ref{tab:2}. We analyzed these and other
dopings and again found universal and parameter-sensitive features.
The parameter-sensitive property is again the presence or absence of
accidental nodes in the $s$-wave gap along the electron FSs. For
most of the parameters, the gap does not have nodes (see Fig.~\ref{fig:2})
because $u_{he}$ increases once it acquires an
additional contribution $u_{h_3 e}$, but for some parameters we still
found nodes along the electron FSs. The universal observations are
that, as long as both hole and electron pockets are present, (i) the
$s$-wave is the leading instability ($\lambda_s > \lambda_d > 0$),
  and (ii) the driving force for the
attraction in both $s$- and $d$- channels is again inter-pocket
electron-hole interaction ($u_{he}$ and ${\tilde u}_{he}$
terms), {\it no matter how small the electron pockets are}.
   In the $d$-wave channel, the electron-hole interaction
changes sign between the two hole FSs
 at $(0,0)$, as a result $d$-wave gaps on these FS have a
$\pi$-phase shift (see Fig.~\ref{fig:2}(c)).

The situation rapidly changes once electron pockets disappear. The
$d$-wave eigenvalue $\lambda_d$ grows relative to $\lambda_s$ and
for the doping shown in Fig.~\ref{fig:2} almost exceeds it. It is
very likely that $d$-wave becomes the leading instability at even
higher dopings, and we therefore focus on the $d$-wave channel.
Comparing ${\tilde u}$ in Table~\ref{tab:2} for the cases with and
without electron pockets, we find that the $d$-wave channel is
attractive in the absence of the electron-hole interaction because of
two reasons. First, the $d$-wave intra-pocket interaction ${\tilde
u}_{h_3 h_3}$ becomes negative (attractive). Second, the
inter-pocket interaction ${\tilde u}_{h_1 h_2}$ is larger in
magnitude than repulsive ${\tilde u}_{h_1 h_1}$ and ${\tilde
u}_{h_2 h_2}$. The solutions with positive $\lambda_d$ then exist
separately for FSs $h_{1,2}$ and $h_3$, and the residual
inter-pocket interaction just sets the relative magnitudes and
phases between the gaps at $h_3$ and $h_{1,2}$. Because ${\tilde
u}_{h_1 h_2}$ is attractive, the two $d$-wave gaps at $h_{1,2}$
are now in phase,
  i.e., this $d$-wave solution is a different eigenfunction
from the one with phase shift $\pi$ at smaller dopings. The
difference is clearly seen by comparing panels (c) and (f) in Fig.~\ref{fig:2}. The $d$-wave gap symmetry at large doping and in-phase
structure of the gaps at $h_{1,2}$ is consistent with the 
%numerical
 fRG solution~\cite{KFeAs_fRG}

{\it Conclusions.} The key result of this work
is the observation that the mechanism of the pairing in FeSCs with hole and electron FSs is different from the one at strong hole or electron doping,
when only one type of FS remains. At small/moderate dopings, the
pairing is driven by inter-pocket electron-hole interaction, no
matter how small hole or electron FSs are. In hole-doped FeSCs, the
leading instability is $s$-wave, while in electron-doped FeSCs, $s$- and
$d$-wave channels are strong competitors, and which of the two wins
depends on the model parameters. At large electron and hole dopings, $d$-wave is the leading
instability, although the $s$-wave channel remains attractive. At
strong electron doping, the origin of the pairing is a direct
$d$-wave attraction between electron pockets. At strong hole doping, however,
the reason for the $d$-wave pairing is a $d$-wave attraction within the
$(\pi,\pi)$ pocket
and between the two hole pockets at $(0,0)$. The $d$-wave pairing at strong hole doping is consistent with the
observation of nodal quasiparticles~\cite{KFeAs_exp_nodal} in
 the heavily hole doped superconductor \KFA with $T_c = 3$K.
Superconductivity at heavy electron
doping at a rather high $T_c \sim 30$K has been recently discovered in \AFS
(A=K, Cs, Rb), which only have electron FSs, according to recent ARPES
studies~\cite{exp:AFESE_ARPES}.
Whether this is a $d$-wave superconductor remains to be seen.

We have only studied the strictly 2D case thus far, and neglected
aspects of the 3D I4/mmm crystal symmetry characteristic of 122
materials and  the hybridization of electron pockets in the folded zone~\cite{mazin}. We nevertheless believe that the general evolution of
interactions and gap symmetry discussed here will be generic to the
 FeSCs.

We acknowledge helpful discussions with L. Benfatto, R. Fernandes,
W. Hanke, I. Eremin, Y. Matsuda, I. Mazin, R.
Prozorov, D. Scalapino, Z.~Tesanovic, R. Thomale, M.
Vavilov, and A. Vorontsov. This work was supported by
NSF-DMR-0906953 (SM and AVC),
 DOE DE-FG02-05ER46236 (PJH),
the Center for Nanophase Materials Sciences, sponsored at ORNL by the Office of Basic Energy Sciences, DOE (TAM), and RFBR 09-02-00127, Presidium of RAS program N5.7, FCP GK P891, and President of Russia MK-1683.2010.2 (MMK). We are grateful to KITP at Santa Barbara for its hospitality during the work on this manuscript.


\begin{thebibliography}{5}
\bibitem{comm_2}
We used the same bandstructure parameters as in
Ref.\protect\cite{graser} and set $\mu = \pm 0.18$  ($n_e=6.22$
and $n_e=5.53$), for the cases of tiny electron or hole FSs, and
$\mu = \pm 0.3$ ($n_e =6.31 $ and $n_e=4.88$) for the cases when
no hole or electron FSs remain  (all parameters are in eV). 
We also did computations for $\mu = \pm 0.05$ and $\mu =\pm 0.1$ and obtained the same results as for $\mu = \pm 0.18$. For electron doping
 we used 6 different set of parameters and got similar results for all sets.
The ones we used in Figs. \protect\ref{fig:1} and \protect\ref{fig:2} are
 $U=1$, $J=0.25$, $V=0.69$.
For hole doping, we used $U=0.9, J=0,
V=0.9$ for $\mu=-0.18$ and  $U=0.75$, $J=0.15$, $V=0.46$ for $\mu=-0.3$, 
 to avoid spin ordering.
\end{thebibliography}
\end{document}